\documentclass[english]{article}
\usepackage[utf8]{inputenc}
\usepackage{amstext}
\usepackage{amssymb}
\usepackage{graphicx}
\usepackage{esint}

\newcommand{\ug}{ \; = \; }

\newcommand{\bb}{\begin{equation}}
\newcommand{\ee}{\end{equation}}
\newcommand{\bega}{\begin{eqnarray}}
\newcommand{\ega}{\end{eqnarray}}
\newcommand{\begae}{\begin{eqnarray*}}
\newcommand{\egae}{\end{eqnarray*}}

\newcommand{\h}{\hspace*{4ex}}

\newcommand{\om}{\omega}

\makeatletter

\newcommand{\address}[1]{
\par {\raggedright #1
\vspace{1.4em}
\noindent\par}
}

\usepackage{cite}

\makeatother

\usepackage{babel}
\begin{document}

\baselineskip 0.55cm

\title{Propagation of time-truncated Airy-type pulses
in media with quadratic and cubic dispersion}

\author{José Angel Borda Hernández$^{1}$, Michel Zamboni-Rached$^{2}$\footnote{On
leave from DECOM-FEEC, University of Campinas}, Amr Shaarawi$^{3}$
and  Ioannis M. Besieris$^{4}$}

\maketitle

\address{$^{1}$FEEC, State University of Campinas, Campinas, SP, Brazil

$^{2}$Department of Electrical and Computer Engineering at the
University of Toronto, Toronto, ON, Canada

$^{3}$Department of Physics, The American University of Cairo, PO
Box 74, New Cairo, Egypt

$^{4}$The Bradley Department of Electrical and Computer
Engineering, Virginia Polytechnic Institute and State University,
Blacksburg VA 24060, USA}

\begin{abstract}
In this paper, we describe analytically the propagation of
Airy-type pulses truncated by a finite-time aperture when second
and third order dispersion effects are considered. The
mathematical method presented here, based on the superposition of
exponentially truncated Airy pulses, is very effective, allowing
us to avoid the use of time-consuming numerical simulations. We
analyze the behavior of the time truncated Ideal-Airy pulse and
also the interesting case of a time truncated Airy pulse with a
``defect'' in its initial profile, which reveals the self-healing
property of this kind of pulse solution.

\end{abstract}

\section{Introduction}

\h For a few years a new kind of localized wave solution has
attracted the attention of researchers. It is the non-dispersive
Airy pulse \cite{Berry1979}. Contrary to other types of
non-dispersive pulses, the Airy one does not require a spectral
space-time coupling to become resistant to the dispersive effects.
Instead, it just uses a cubic phase frequency spectrum, which
greatly facilitates its experimental generation.

\h An ideal Airy pulse is immune to dispersion effects of second
order, but it possesses infinite energy making its experimental
generation impossible. However, it is possible to obtain a
finite-energy version of the ideal Airy pulse by modulating it
with a time exponential function on the initial plane $z=0$
\cite{Siviloglou2007a,Besieris2008}. The analysis of a
finite-energy Airy pulse considering the effects of second and
third order dispersion was first made in \cite{Besieris2008} and
subsequently in \cite{Cai2013,Driben2013,Driben2013a}. Additional
work on finite-energy Airy beams and pulses is provided in
Refs.\cite{Abdollahpour2010,Ament2011,Ren2013,Nerukh2011,Couairon}.

\h All the previous analyses about finite-energy Airy pulses have
considered exponential or Gaussian (initial) time apodization, but
until now no one has studied the case of Airy-type pulses with
truncation in time, that is, a temporal apodization made with a
rectangular function. Such truncation, besides providing finite
energy solutions, enables the resulting pulse to present, up to a
certain distance, the very same characteristics of the
non-truncated version. Other types of apodizations, such as the
exponential and Gaussian, are not so efficient in this sense.

\h In this paper, we present an analytical description of
Airy-type pulses truncated in time when second and third order
dispersion effects are considered. The mathematical method used
here is an extension of one used for describing spatially
truncated Airy beams \cite{Zamboni-Rached2012}.

\h We analyze the behavior of the time truncated Ideal-Airy pulse
and also the interesting case of a time truncated Airy pulse with
a ``defect'' in its initial profile, which reveals the
self-healing property of this kind of pulse solution.

\section{Theoretical description of Airy-type pulses truncated in time}

\h When quadratic and cubic dispersion effects are predominant in
the evolution of a pulse in a linear and homogeneous material
medium, the differential equation describing the pulse envelope
propagation can be written as \cite{Agrawal2012}

\begin{equation}
i\frac{\partial U}{\partial
z}=\frac{\beta_{2}}{2}\frac{\partial^{2}U}{\partial T^{2}}+
\frac{i\beta_{3}}{6}\frac{\partial^{3}U}{\partial
T^{3}}\label{eqdifparcial}
\end{equation}

where $T$ is the retarded time in a frame reference moving with
the group velocity ($T=t-z/v_{g}$), and $\beta_{2}$ and
$\beta_{3}$ represent the second and third-order dispersion
parameters, respectively, both depending on the central frequency
of the pulse.

\h Let us define $b=-\beta_{3}/\left(2\beta_{2}T_{0}\right)$ and
the new normalized variables $s=T/T_{0}$ and
$Z=z\beta_{2}/T_{0}^{2}$, where $T_0$ is a constant that will be
related with the time width of the initial pulse.

\h Now, by considering an initial temporal pulse profile given by
an Airy function apodized by an exponential, i.e,

\begin{equation}
U_{n}(s,Z=0)=\textrm{Ai}\left(s\right)\exp\left(a_{n}s\right)
 \,\, , \label{pulsoinicial1}
\end{equation}

with $Re(a_n)>0$, we can obtain, by solving
eq.(\ref{eqdifparcial}), the finite energy Airy pulse given by

\begin{eqnarray}
U_{n}\left(s,Z\right) & = & \frac{1}{\left(1+bZ\right)^{1/3}}\nonumber \\
 & \times & \exp\left[\frac{6a_{n}\left(2s-Z^{2}\right)+iZ\left(-6a_{n}^{2}-6s+Z^{2}\right)}{12\left(1+bZ\right)^{2}}\right]\nonumber \\
 & \times & \exp\left[b\frac{4Za_{n}\left(-a_{n}^{2}+3s+a_{n}^{2}bZ\right)+i6Z^{2}\left(a_{n}^{2}-s\right)}{12\left(1+bZ\right)^{2}}\right]\nonumber \\
 & \times & \mathrm{Ai}\left[\frac{\left(s-\frac{Z^{2}}{4}-ia_{n}Z\right)+bZ\left(s-
 a_{n}^{2}\right)}{\left(1+bZ\right)^{4/3}}\right]\,\, , \label{airyfinito}
\end{eqnarray}

which was first obtained and analyzed in detail by Besieris
\emph{et al.} in \cite{Besieris2008}. Briefly, when $a_n=0$
(infinite energy), $\beta_3=0$ and $\beta_2 \neq 0$, we have the
Ideal-Airy pulse, which is immune to the dispersion effects. In
the case $a_n=0$, $\beta_3 \neq 0$ and $\beta_2 \neq 0$, the
Ideal-Airy pulse can present resistance to the dispersion effects
for long (finite) distances.

Now, when $Re(a_n)>0$ the corresponding pulse given in
eq.(\ref{airyfinito}) possesses finite energy content and it is
dispersion resistant for long distances when $Re(a_n)>> 1/T_0$.
However, even in these cases, the entire pulse starts to be
destorted from the beginning and this occurs because with the
exponential apodization at $z=0$, all time sidelobes are affected
(damped) and, as it is well known, the latter are responsible for
the pulse reconstruction.

\h Based on these observations, we can envisage that a time
apodization at $z=0$ made by a rectangular function (i.e., a
double step function) will preserve, within it, exactly the same
initial field structure of the non-apodized pulse. As a result, we can
expect that the  resulting pulse will maintain the very same
characteristics of its non-truncated version if the time width
of the truncation is much greater than $T_0$. Actually, in the
self-healing process, the inner lobes (i.e., those closer to
the main peak) are fed by the outer ones. The latter are consummed
until only the main peak remains, which then suffers distortions
due to the dispersion effects. At this point, the
pulse will have reached its maximum distance of resistance from the
dispersion effects (depth of field).

\h Next, we are going to present a mathematical method capable of
describing Airy-type pulses apodized by a time rectangular
function i.e., truncated in time, propagating in a linear medium
characterized by both quadratic and cubic dispersion. This
approach is a variant/extension of one developed to deal with
spatially truncated Airy beams\cite{Zamboni-Rached2012}, where the
equivalent to the third order term in eq.(\ref{eqdifparcial}) was
not considered.

\h Mathematically, we wish to solve (approximately)
eq.(\ref{eqdifparcial}), when the following
initial pulse profile is considered at $z=0$

\bb F(s) \ug \textrm{Ai}\left(s\right)m\left(s\right)
\left[u\left(s+S\right)-u\left(s-S\right)\right] 
\label{pulsotruncadoinicial}\ee

\h Here, $m(s)$ is an arbitrary function modulating the Airy
function and the time truncation is represented by the difference
between the Heaviside unit step functions $u(s+S)$ and $u(s-S)$.

\h We start by considering as a solution to
eq.(\ref{eqdifparcial}) a superposition of pulses given by

\begin{eqnarray}
U\left(s,Z\right) & = & \sum_{n=-\infty}^{\infty} B_n
U_{n}\left(s,Z\right) \,\, , \label{airytruncado}
\end{eqnarray}

where the pulses $U_n$ are given by (\ref{airyfinito}), with $B_n$
and $a_n$ complex constants yet unknown.

\h The pulse solution (\ref{airytruncado}) at $z=0$ is written as

\begin{equation}
U\left(s,0\right)=\sum_{n=-\infty}^{\infty}B_n
U_{n}\left(s,0\right)=
\sum_{n=-\infty}^{\infty}B_{n}Ai\left(s\right)\exp\left(a_{n}s\right)\label{airytruncadoinicial}
\end{equation}

Our problem is to determine the values of $B_n$ and $a_n$ so that
our proposed solution at $z=0$ is (approximately) equal to the
desired truncated Airy-type pattern given by
eq.(\ref{pulsotruncadoinicial}). Once this is done, the resulting
pulse propagating in the second and third-order dispersion medium
is given by eq.(\ref{airytruncado}).

\h Let us make the following choice:

\begin{eqnarray}
a_{n} & = & a_{R}+i\frac{2\pi}{L}n\label{an}
\end{eqnarray}

where $a_{R}$ and $L$ are positive constants. By using (\ref{an})
in (\ref{airytruncadoinicial}), we get

\begin{eqnarray}
U\left(s,0\right) & = &
\textrm{Ai}\left(s\right)\exp\left(a_{R}s\right)
\sum_{n=-\infty}^{\infty}B_{n}\exp\left(i\frac{2\pi}{L}ns\right)\label{pulsotruncadoinicial2}
\end{eqnarray}

\h The Fourier series appearing in
eq.(\ref{pulsotruncadoinicial2}) is suggestive. In the case we
define

\begin{eqnarray}
\Lambda\left(s\right) & \equiv &
\sum_{n=-\infty}^{\infty}B_{n}\exp\left(i\frac{2\pi}{L}ns\right)
\,\, , \label{Lamb}
\end{eqnarray}

with $S < L/2$, and the coefficients $B_{n}$ as

\begin{eqnarray}
B_{n} & = & \frac{1}{L}\int_{-S}^{S}m\left(s\right)
\exp\left(-a_{R}s\right)\exp\left(-i\frac{2\pi}{L}ns\right)ds \,\,
, \label{Bn}
\end{eqnarray}

the series (\ref{Lamb}) will represent, within the domain $-L/2\leq s\leq
L/2$, the following function

\begin{equation}
\Lambda\left(s\right)=\left\{ \begin{array}{ccc}
m\left(s\right)\exp\left(-a_{R}s\right) & \textrm{for} & -S\leq s\leq S\\
0 & \textrm{for} & S < \left|s\right|\leq L/2
\end{array}\right.\label{Lamb2}
\end{equation}

\h In this way, by using
(\ref{pulsotruncadoinicial2},\ref{Lamb},\ref{Bn},\ref{Lamb2}) and
with appropriate values for $a_R$ and $L$, we can get the following
result:

\begin{equation}
U\left(s,Z=0\right)=\left\{ \begin{array}{ccc}
\textrm{Ai}\left(s\right)m\left(s\right) & \textrm{for} & \left|s\right|\leq S\\
0 & \textrm{for} & S < \left|s\right|\leq L/2\\
Ai\left(s\right)\exp\left(a_{R}s\right)\Lambda\left(s\right)\thickapprox0
& \textrm{for} & \left|s\right|>L/2
\end{array}\right. \,\, \approx \,\, F(s) \label{airytruncadoinicial2}
\end{equation}

The suitable values for $L$ and $a_R$ serve to guarantee the
result given above when $|s| > L/2$. Actually, as $\Lambda(s)$ is
the Fourier series with period $L$ representing the function given
in eq.(\ref{Lamb2}) and since $L/2>S$, \emph{for appropriate
choices} of $L$ and $a_R$ we have that $Ai(s)
\exp(a_R\,s)\Lambda(s)\approx 0$ for $|s| > L/2$ due to the
behavior of the functions $Ai(s)$ and $\exp(a_R s)$ for positive
and negative values of $s$, respectively. A good criterion is to
choose values of $a_R$ and $L$ such that\footnote{Of course there
are many sets of values of $a_R$ and $L$ that satisfy this
criterion, yielding a good result.} $\exp(a_R L/2)
>> \rm{Max}[\Lambda(s)]$, with $\Lambda(s)$ given in Eq. (\ref{Lamb2}).

\h So, we have achieved our goal: The solution describing the
propagation of an Airy-type pulse truncated in time according to
eq.(\ref{airytruncadoinicial2}) in the presence of quadratic and
cubic dispersion is given by eq.(\ref{airytruncado}), with $a_n$
and $B_n$ given by eqs.(\ref{an},\ref{Bn}), $a_R$ and $L$ being
appropriately chosen according the above criterion.

\h We can also calculate the initial temporal spectrum of the
time-truncated Airy-type pulse. Fourier transformation of
(\ref{airytruncadoinicial}) yields

\begin{equation}
\small\widetilde{U}\left(0,\Omega\right)=
\sum_{n=-\infty}^{\infty}B_{n}\exp\left(-a_{n}\Omega^{2}\right)
\exp\left(\frac{i}{3}\left(\Omega^{3}-3a_{n}^{2}\Omega-ia_{n}^{3}\right)\right)\,
. \label{spec}
\end{equation}

\textbf{Examples}

\h Here, we are going to apply the method considering two
situations where the material medium is fused silica, the central
wavelength $\lambda_{0}=1550$nm (so $\beta_2=-27.909 \rm{ps}^2/$km
and $\beta_3=0.151 \rm{ps}^3/$km), $T_{0}=10$ps, $a_{R}=0.1$,
$S=40$ and $L=3S$. Of course a finite number of $2N+1$ terms must
be used in the summation of our solution given by
eq.(\ref{airytruncado}), corresponding to $-N\leq n\leq N$. Here
we use $N=80$.

\

1. \emph{Ideal Airy pulse truncated in time}

%
%


\h Let us consider the case of an Ideal-Airy pulse with initial
main peak of $T_0 = 10$ps, truncated by a finite time aperture of
width forty times greater than $T_0$, i.e., $S=40$. In this case
we have $m(s)=1$ and the pulse at $z=0$ is given by
eq.(\ref{airytruncadoinicial2}), where the coefficients $a_{n}$
and $B_{n}$ are defined by the equations (\ref{an}) and
(\ref{Bn}), respectively. As already has been mentioned, many sets
of values for $a_R$ and $L$ can yield good results and here we
have chosen $L=3S$ and $a_{R}=0.1$. The resulting pulse is given
by eq.(\ref{airytruncado}).

\begin{figure}[!h]
\begin{center}
\scalebox{1.3}{\includegraphics{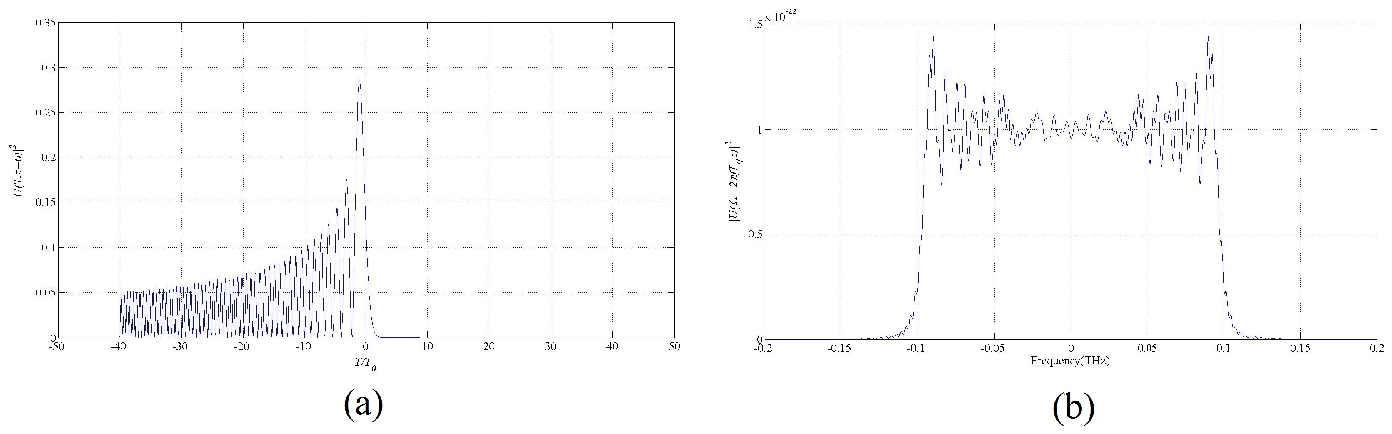}}
\end{center}
\caption{(a)Field intensity, at $z=0$, of the Ideal-Airy pulse
truncated by a finite-time aperture according to example 1; (b)
Frequency spectrum of the truncated Ideal Airy pulse} \label{fig1}
\end{figure}

\h Figure \ref{fig1}a shows the pulse intensity at $z=0$ obtained
from equation (\ref{airytruncadoinicial}), and Figure \ref{fig1}b
shows its frequency spectrum given by eq.(\ref{spec}).

\newpage

\begin{figure}[!h]
\begin{center}
\scalebox{2}{\includegraphics{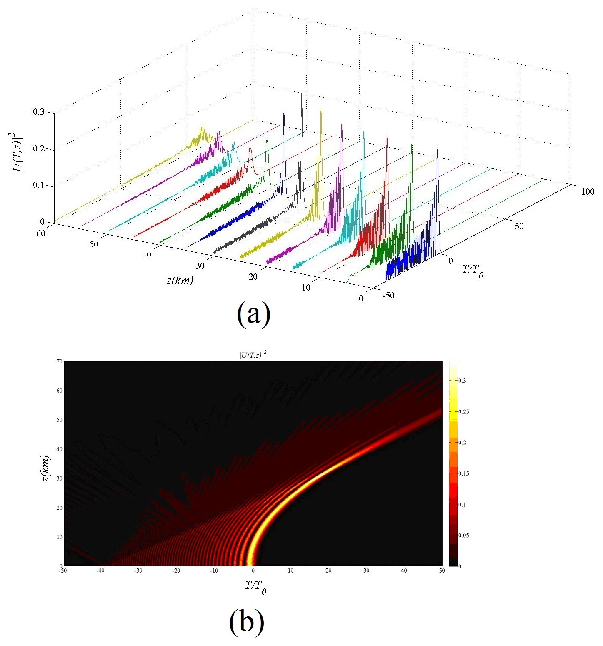}}
\end{center}
\caption{(a)Temporal evolution of the time-truncated Ideal Airy
pulse intensity of example 1 at different propagation distances;
(b) The same pulse evolution through an orthogonal projection on
the plane $(T/T_0,z)$.} \label{fig2}
\end{figure}




\newpage

\h The temporal evolution of the resulting pulse at different
distances is shown in Figure \ref{fig2}a. We can see that with
this kind of time apodization the inner lobes, in particular the
main one, are fed by the outer ones and maintain their
amplitude and shape similarly to those of the Ideal-Airy pulse. This
process continues until the distance (field depth) where only the
main peak remains, which then is distorted by
dispersion effects. We should note that at that distance the pulse
has a shape that is very different from the initial one, given
that there are no longer any sidelobes left, but just the main peak with
the same time width $T_0$.

\h For comparison, a Gaussian pulse with the same
initial time and central wavelength,
would have a dispersion length $L_D = T_0^2/|\beta_2| \approx
3.6$km, while the field depth of the present truncated Ideal Airy
pulse is approximately $35$km.

\h Figure \ref{fig2}b shows the same pulse evolution through an
orthogonal projection on the plane $(T/T_0,z)$, where
the accelerated character of the pulse is evident.

\begin{figure}
\begin{centering}
\includegraphics[scale=1.2]{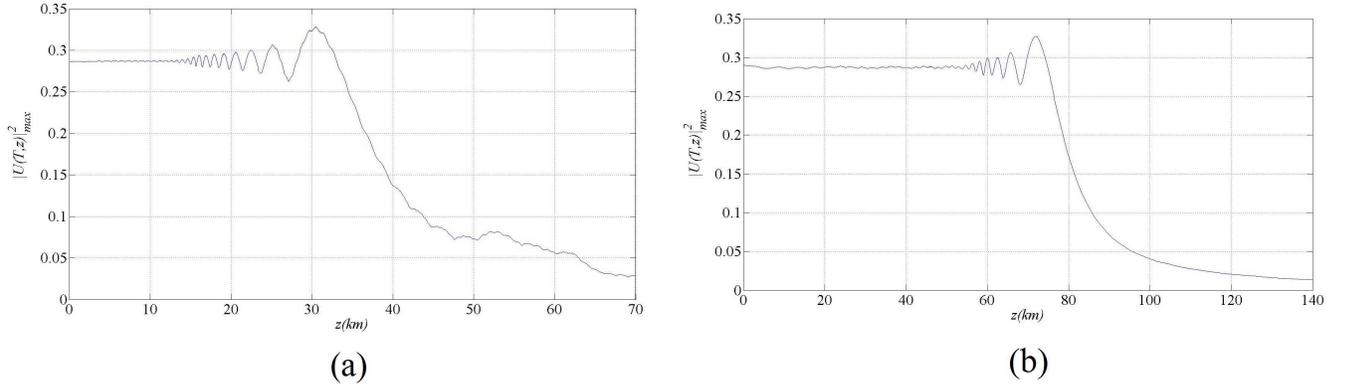}
\par\end{centering}
\caption{Peak intensity evolution in range when $S =
T/T_0 = 40$ and $150$ respectively.}\label{fig3}
\end{figure}

\h It should be clear that the greater the time truncation width,
the greater the depth of field of the resulting pulse. Figures
\ref{fig3}a and \ref{fig3}b show the peak intensity evolution
along the Z-direction when $S = T/T_0 = 40$ and\footnote{In the
case of $S=150$ we use $a_R=0.01$.} $150$, respectively. Of
course, a wider time width requires a larger amount of energy.

%





\h Many other Airy-type pulses truncated in time can be described
through the present method just by choosing the appropriate
modulation function $m(s)$ in eq.(\ref{airytruncadoinicial2}).

\

2. \emph{Self reconstruction of a defective Airy pulse truncated
in time}

\h Here, we are going to present the analytical description of the
evolution of a time truncated Airy pulse with a ``defect'' in its
initial profile. This situation can be easily described by our
method, since the defect in question can be inserted by the
function $m(s)$ in the initial pulse profile,
eq.(\ref{airytruncadoinicial2}). For instance, let us consider the
Truncated Ideal Airy pulse of the previous example with a missing
piece of its initial field. This defect can be represented by
choosing

\begin{equation}
m\left(s\right)=\left\{ \begin{array}{ccc}
0 & \textrm{for} & -S_2\leq s\leq -S_1\\
1 & \textrm{elsewhere}
\end{array}\right.\label{m}
\end{equation}

\newpage

\begin{figure}[!h]
\begin{center}
\scalebox{1.3}{\includegraphics{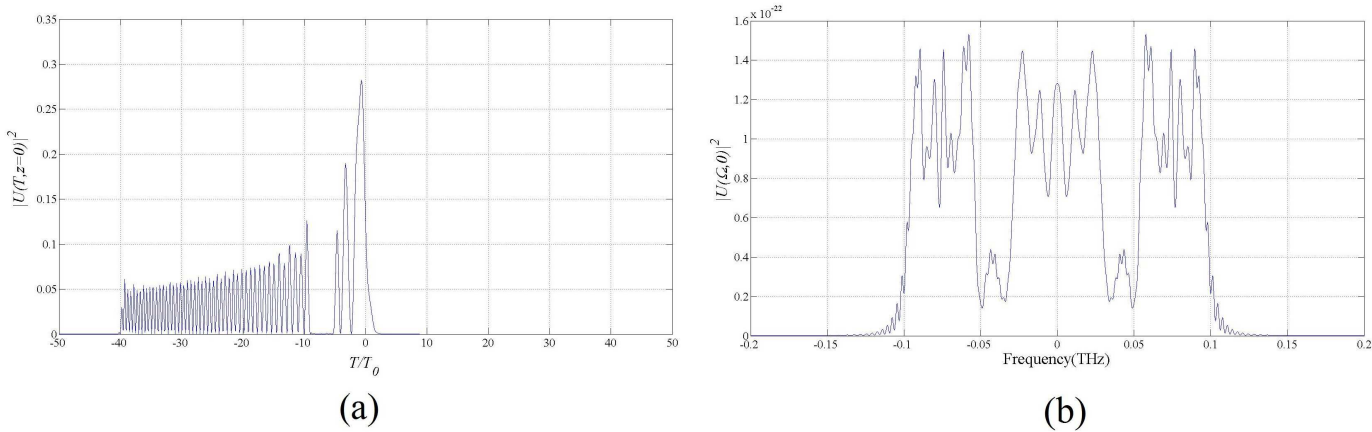}}
\end{center}
\caption{(a)Field intensity, at $z=0$, of a defective Airy pulse
truncated by a finite-time aperture; (b) Frequency spectrum of the
truncated defective Airy pulse} \label{fig4}
\end{figure}

\begin{figure}[!h]
\begin{center}
\scalebox{2}{\includegraphics{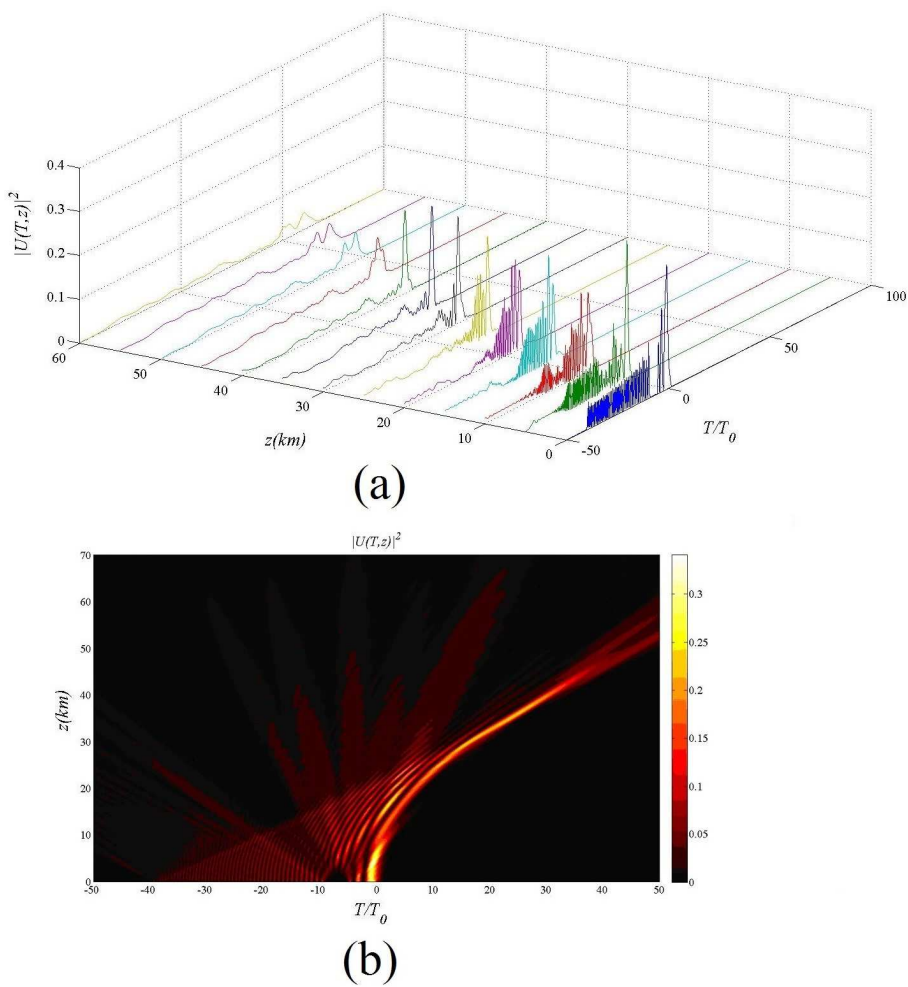}}
\end{center}
\caption{(a)Temporal evolution revealing the self-healing property
of a defective Airy pulse truncated by a finite-time aperture; (b)
The same pulse evolution through an orthogonal projection on the
plane $(T/T_0,z)$.} \label{fig5}
\end{figure}

\newpage

\h If this ``gap'' in the pulse profile is considerable shorter
than the truncation time width, the Airy-type pulse is able to
self-regenerate. To confirm this important property we set $S_1 =
5$ and $S_2 = 9$, such that the defect occupies $10\%$ of the
initial pulse.

\h This defect can be clearly seen in Fig.\ref{fig4}a, which shows
the initial ($z=0$) pulse intensity, obtained from
eq.(\ref{airytruncadoinicial}). Figure \ref{fig4}b shows the
correspondent frequency spectrum, given by eq.(\ref{spec}).

\h The temporal evolution of the resulting pulse at different
distances is shown in Fig.\ref{fig4}a, where the self-healing
property is revealed. The main part of the pulse (close to the
main peak) is severely affected in the range $10 \rm{km}  \lesssim
z \lesssim  25$km, after which the main peak is regenerated,
lasting until $ z \approx 35$km, when it gets to be distorted by
dispersion effects.

\h Figure \ref{fig5}b shows the same pulse evolution through an
orthogonal projection on the plane $(T/T_0,z)$. Again, the
accelerated character of the pulse is evident.

\section{Conclusion}

\h Contrary to previous studies of finite-energy Airy beams and
pulses, as well as of other types of localized waves
\cite{Besieris96,Besieris98}, based on exponential or Gaussian
spatial or temporal apodization techniques, a novel efficient
time-truncation method has been developed in this article and has
been used to examine the evolution of a time-truncated ideal Airy
pulse in a homogeneous linear medium characterized by both second
and third-order temporal dispersion. A detailed theoretical study
has been undertaken of the depth of field, and illustrations have
been provided clearly showing the acceleration property of such a
pulse. It has been established that a properly truncated Airy
pulse can maintain its salient characteristic features up to
ranges that are much larger than those of a Gaussian pulse with a
comparable initial spectral structure. The same method has been
used to study the evolution of a "deficient" ideal Airy pulse in
the same medium, illustrating the accompanying regeneration or
self-healing effects.

\h The work of Chong \emph{et al}. \cite{Chong2010} on versatile
light bullets limited to second-order temporal dispersion can be
extended to accommodate both second and third-order dispersive
effects. Eq.(\ref{eqdifparcial}) is extended as follows:

\begin{equation}
i\frac{\partial \psi(\textbf{r},t)}{\partial z} \ug
\frac{\beta_{2}}{2}\frac{\partial^{2}\psi(\textbf{r},t)}{\partial
T^{2}}+
\frac{i\beta_{3}}{6}\frac{\partial^{3}\psi(\textbf{r},t)}{\partial
T^{3}} - \frac{1}{2\beta_0}\nabla^2_t\psi(\textbf{r},t)
\label{eq3D}
\end{equation}

\h Here, $\nabla^2_t$ denotes the transverse (with respect to $z$)
Laplacian operator and $\beta_0$ is the wavenumber computed at the
central angular frequency $\om_0$. Equation (\ref{eq3D}) allows a
solution of the form

\bb \psi(\textbf{r},t) \ug U(z,T)Q(\textbf{r}) \label{UQ} \ee

with $U(z,T)$ satisfying Eq.(\ref{eqdifparcial}) and
$Q(\textbf{r})$ governed by the 3D parabolic equation

\begin{equation}
i\frac{\partial Q(\textbf{r})}{\partial z} +
\frac{1}{2\beta_0}\nabla^2_t Q(\textbf{r}) \ug 0 \label{Q}
\end{equation}

\h A time-truncated ideal Airy beam solution $U(z,T)$ together
with a finite-energy solution of the paraxial equation (\ref{Q}),
e.g., a Bessel-Gauss beam, will give rise to a light bullet
$\psi(\textbf{r},t)$ according to Eq.(\ref{UQ}). In this case, the
finite energy light bullet can be resistant to the diffraction and
dispersion effects for a distance much greater than the field
depth of the ordinary pulses in dispersive media. If an Airy beam
solution to Eq.(\ref{Q}) is chosen, a nonlinear transverse bending
due to diffraction will appear in addition to the longitudinal
acceleration.

\h This work was supported by FAPESP (under grant 2013/26437-6);
CNPq (under grants 312376/2013-8) and CAPES.

\h The authors thank Erasmo Recami and Mo Mojahedi for valuable
discussions and kind collaboration.

\end{document}